\documentclass[acmsmall]{acmart}

\AtBeginDocument{%
  }

\setcopyright{acmlicensed}
\copyrightyear{2018}
\acmYear{2018}
\acmDOI{XXXXXXX.XXXXXXX}

\acmJournal{JACM}
\acmVolume{37}
\acmNumber{4}
\acmArticle{111}
\acmMonth{8}

\usepackage{multirow}
\usepackage{xspace}
\usepackage[ruled,linesnumbered]{algorithm2e}

\newcommand{\modelname}{CDCDA-PLM\xspace}

\begin{document}

\title[Towards On-Device Personalization]{Towards On-Device Personalization: Cloud-device Collaborative Data Augmentation for Efficient On-device Language Model}

\author{Zhaofeng Zhong}
%\orcid{0000-0003-3493-6985}
\affiliation{%
  \institution{The University of Queensland}
  \city{Brisbane}
  \country{Australia}
}
\email{zhaofeng.zhong@uq.edu.au}

\author{Wei Yuan}
% \authornote{Corresponding authors.}
\affiliation{%
  \institution{The University of Queensland}
  \city{Brisbane}
  \country{Australia}
}
\email{w.yuan@uq.edu.au}

\author{Liang Qu}
\affiliation{%
  \institution{The University of Queensland}
  \city{Brisbane}
  \country{Australia}
}
\email{liang.qu@uq.edu.au}

\author{Tong Chen}
\affiliation{%
  \institution{The University of Queensland}
  \city{Brisbane}
  \country{Australia}
}
\email{tong.chen@uq.edu.au}

\author{Hao Wang}
\affiliation{%
  \institution{Alibaba Cloud Intelligence Group}
  \country{China}
}
\email{cashenry@126.com}

\author{Xiangyu Zhao}
\affiliation{%
  \institution{City University of Hong Kong}
  \city{Hong Kong}
  \country{China}}
\email{xianzhao@cityu.edu.hk}

\author{Hongzhi Yin}
% \authornote{Corresponding authors.}
\affiliation{%
  \institution{The University of Queensland}
  \city{Brisbane}
  \country{Australia}
}
\email{h.yin1@uq.edu.au}
\thanks{Corresponding Authors: Wei Yuan and Hongzhi Yin}
\renewcommand{\shortauthors}{Zhong et al.}

\begin{abstract}
  % LLMs 
With the advancement of large language models (LLMs), significant progress has been achieved in various Natural Language Processing (NLP) tasks. 
However, existing LLMs still face two major challenges that hinder their broader adoption: (1) their responses tend to be generic and lack personalization tailored to individual users, and (2) they rely heavily on cloud infrastructure due to intensive computational requirements, leading to stable network dependency and response delay.
Recent research has predominantly focused on either developing cloud-based personalized LLMs or exploring the on-device deployment of general-purpose LLMs. However, few studies have addressed both limitations simultaneously by investigating personalized on-device language models.
To bridge this gap, we propose \modelname, a framework for deploying personalized on-device language models on user devices with support from a powerful cloud-based LLM. 
Specifically, \modelname leverages the server-side LLM's strong generalization capabilities to augment users' limited personal data, mitigating the issue of data scarcity. 
Using both real and synthetic data, a personalized on-device language model (LM) is fine-tuned via parameter-efficient fine-tuning (PEFT) modules and deployed on users' local devices, enabling them to process queries without depending on cloud-based LLMs. 
This approach eliminates reliance on network stability and ensures high response speeds.
Experimental results across six NLP personalization tasks demonstrate the effectiveness of \modelname.

\end{abstract}




\received{20 February 2007}
\received[revised]{12 March 2009}
\received[accepted]{5 June 2009}

%% remove ACM reference
% \settopmatter{printacmref=false} %remove ACM reference format

\maketitle

\section{Introduction}
Recently, Large Language Models (LLMs) have become a cornerstone of contemporary Natural Language Processing (NLP) research and industry applications due to their exceptional abilities in text understanding and generation~\cite{radford2018improving,ray2023chatgpt,naveed2023comprehensive}. 
These models have achieved remarkable success and transformed numerous areas of NLP, such as translation, summarization, and conversational AI~\cite{thirunavukarasu2023large,hu-etal-2024-gentranslate,wang2024survey}. 

Despite their advancements, existing LLMs face two significant limitations that hinder their broader adoption:
(1) \textbf{Lack of Personalization}. LLMs are designed as universal models, which limits their ability to generate responses tailored to users' personalized preferences and interests;
(2) \textbf{Dependence on Cloud Infrastructure}. The powerful LLMs are typically trained and deployed on cloud servers due to their high computational demands. 
This architecture not only relies on stable and high-speed network connections to transmit user queries and deliver responses, but also requires a long time for LLM inference. 
However, these conditions are often unmet in real-world scenarios, particularly on mobile platforms such as smartphones and smart vehicles \cite{On-deviceRecKD-xia-SIGIR2022,OnDeviceRec-Survey-yin-2025}.
For instance, LLM-based assistants are increasingly integrated into smart vehicle systems, but these vehicles frequently experience inconsistent network connectivity due to their mobility. 
In remote regions with weak or no network coverage, such cloud-based LLM services could become entirely inaccessible as the system goes offline.
In addition, considering the time-sensitive applications, an LLM-powered service will lead to a significant inference latency overhead in reality, limiting the feasibility of the service.
Consequently, there is a growing need for personalized LLMs that can run directly on user devices. 
Such models must address user-specific needs while operating efficiently on edge devices, free from the constraints of cloud connectivity \cite{LLRec-POIRecOnDevice-wang-www2020,PEEL-EmbeddingLearning-zheng-TKDE2024}.

Some recent efforts have explored techniques for enabling personalization in LLMs, which can be generally categorized into prompt-based methods and fine-tuning-based methods.
The prompt-based approaches format personalized prompts to leverage the in-context learning capabilities of LLMs. 
That is to say, all users share the same model, but personalized prompts are used to guide the generation process.
For example,~\citet{christakopoulou2023userlongcontextperson} incorporates users' historical data into prompts to enhance generation performance. 
And to conquer the input length limitation when users' historical data are too long, some research employs retrieval-augmentation generation (RAG) to augment user's query by adding the most relevant history information into prompt~\cite{richardson2023summaryretrieval,salemi2023lamp,automaticprompt2024,li2024Personalizedhumanfeedback}.
On the other hand, fine-tuning methods directly optimize the parameters of LLMs to adapt to users’ personalized data distributions~\cite{tan2024OPPUPEFTperuser,tan2024personalizedpiecesPEFT,park2024RLHFfeedbackpersonalization,li2024Personalizedhumanfeedback,Zhuang2024HYDRA}. 
However, as the number of users grows, these approaches face significant scalability challenges because the server’s computational cost increases at least linearly with the user base, making large-scale deployment impractical \cite{DCPR-DiffusionPOI-long2024,DARD-CollaborativePOIRec-zheng-www2024}.
Specifically, since a separate personalized model must be trained for each user, a server-based design would require training N models for N users. 
As the number of users grows, this results in a significant computational burden on the server, making the fine-tuning-based personalization method impractical \cite{Efficient-device-session-Rec-xia-tois2023,CommunicateEfficientUpdateRec-xia-CIKM2023,FELLAS-FederatedRecLLM-Yuan-TOIS2025}.

While these methods show promise for personalization, they are primarily designed for cloud-based LLMs and face significant challenges in on-device settings. 
On-device language models (LMs) are constrained by the computational and storage limitations of edge devices, resulting in small model sizes. 
As demonstrated in many previous works \cite{richardson2023summaryretrieval,salemi2024optimization}, prompt-based personalization methods, including RAGs, cannot achieve satisfactory performance with these small-sized on-device LMs since these models have limited generalization and contextual understanding ability.
Similarly, fine-tuning on-device LMs to adapt to users' local data distributions presents additional difficulties. 
For example, individual users typically possess limited data, which is insufficient for effective model fine-tuning. 

In this paper, we take a first step toward developing a learning framework for personalized on-device language models (LMs). 
Our approach leverages cloud-based large language models (LLMs) to address the challenge of personal data scarcity and introduces a cloud-device collaborative framework to ensure scalability. 
Specifically, the cloud-based LLM generates synthetic data tailored to each user's limited local data, thereby augmenting the user’s dataset and transferring relevant knowledge from the cloud LLM to the on-device LM. 
Once users receive the synthetic data, we apply parameter-efficient fine-tuning (PEFT) techniques to optimize their on-device LMs using both the synthetic and local personal data entirely on the user’s device. 
This decentralized training strategy avoids the scalability bottleneck of requiring the cloud server to fine-tune models for all users, parallelized personalization across a large user base.
After training, the personalized on-device LM is deployed locally, allowing inference to be performed without network connectivity, which reduces both network dependency and inference latency. 
To evaluate the framework’s effectiveness, we conduct extensive experiments on public datasets, and experimental results demonstrate that the proposed method can achieve promising performance in personalized classification and generation tasks.

Overall, the main contributions of this paper are summarized as follows:
\begin{itemize}
  \item We take the first step in exploring the problem of LLM personalization in the context of small on-device LM deployment, where storage size and computational resources are constrained.
  \item We propose a personalized on-device LM framework, \modelname. 
  In this framework, we design a novel cloud-device collaboration mechanism in which the server model leverages data augmentation to transfer knowledge to the small on-device LM. 
  Additionally, we develop a dedicated filtering method to enhance the robustness of the knowledge transfer process.
  \item We conduct extensive experiments across multiple tasks to demonstrate the effectiveness of \modelname.
  Furthermore, we perform detailed ablation studies and hyperparameter analyses, followed by a case study, to further highlight the superiority of our proposed method.
\end{itemize}

\section{Related Work}
In this section, we briefly review the literature on LLM personalization, on-device deployment of LLMs and textual data augmentation.

\subsection{Personalization of LLMs}
Personalized LLM aims to better understand and generate text specific to match the user's interests and preferences.
The existing research on LLM personalization could generally be divided into two categories: prompt design based personalization and parameter-efficient fine-tuning (PEFT) based personalization \citep{salemi2024comparingretrievalPEFT}. 

\textbf{Prompt-based Personalization.}
In the early development of personalized prompts, query prompts were formatted with user history as context to leverage the in-context and few-shot learning capabilities of large language models (LLMs).
For instance, \citet{christakopoulou2023userlongcontextperson} and \citet{zhiyuli2023bookgptgeneralframeworkbook} demonstrate that incorporating long user history in prompts can enhance LLM generation performance.
However, incorporating user history in prompts will increase the inference computational cost due to the lengthy input.
To mitigate this issue, \citet{salemi2023lamp} proposed a strategy to shorten the user history length by using a retrieval model to select relevant documents from the user history based on the user query. 
Subsequently, \citet{salemi2024optimization} trained a retriever that selects relevant documents from user history based on the user query, using LLM feedback to refine the retriever and capture more personalized features.
Moreover, \citet{richardson2023summaryretrieval} employed LLMs to generate concise summaries of user history, potentially capturing a more comprehensive perspective of the user.

\textbf{Fine-tuning Based Personalization.}
Parameter-efficient fine-tuning (PEFT) offers an effective way to optimize LLMs for users' personal distributions by modifying only a small subset of parameters \citep{hu2021lora, dettmers2023qloraefficientfinetuningquantized}.
For example, OPPU proposed a PEFT-based personalized LLM, which fine-tunes the LoRA adapter on user profiles for each user, to store user knowledge on PEFT parameters \citep{tan2024OPPUPEFTperuser}.
Building on this work, PER-PCS aggregates fine-tuned LoRA adapters from multiple users into a shared adapter pool, which can be leveraged to generate a personalized LLM for a target user by merging multiple LoRA adapters \citep{tan2024personalizedpiecesPEFT}.
Reinforcement learning is also applied with PEFT to achieve better performance\citep{cheng2023personalizedrewardmodels,li2024Personalizedhumanfeedback, park2024RLHFfeedbackpersonalization}.

However, all the aforementioned personalization approaches have been developed for cloud-based LLMs, which possess formidable generalization and language understanding capabilities, lacking the exploration of weak on-device models.

\subsection{On-device Deployment of LLMs}
Due to their large size, deploying LLMs on edge devices presents critical challenges, including high computational overhead and significant memory demands. Current deployment methods can generally be categorized into two strategies.

The first strategy involves directly compressing the original large-scale model into a smaller one through quantization \cite{liu2023LLMQAT, Lin2025AWQquantization} and pruning \cite{ma2023llmpruner, Frantar2023SparseGPT}. Quantization maps high-precision values to lower precision, while pruning removes certain unimportant neurons. However, since the compressed model remains architecturally coupled with the original model, aggressive compression may lead to significant performance degradation.

The second strategy focuses on transferring knowledge from a large cloud-based model to a smaller on-device model. A widely used approach within this strategy is knowledge distillation (KD) \cite{hinton2015distillingknowledgeneuralnetwork, Gou_2021KnowledgeDistillationSurvey}. 
Based on the accessibility of the teacher model, the KD process can be classified into white-box KD and black-box KD.
In white-box KD, the student model learns from the teacher model’s activations, hidden features, and output distribution \cite{xu2024surveyLLMKD, ko2024distillLLM, wu2024adaptiveKLdivergence, agarwal2024JSDivergence, gu2024minillmknowledgedistillationlarge}. However, this approach requires the student model to share certain architectural similarities with the teacher model.
In contrast, black-box KD allows the student model to access only the teacher model’s responses to enhance training data \cite{Dai2025AugGPT,ho2023llmAreReasoningTeacher, tian2024studentfrommultipleteacherLLM, jung-etal-2024-impossibleDistill}.
For instance, \citet{Qin2024ondeviceDataSelection} introduces an on-device LLM training framework by selecting the most representative user data to mitigate the data storage demands in the device. 
However, in their method, the on-device model is as large as the cloud-based model, which is impractical.
Our proposed method aligns closely with black-box KD, leveraging a cloud-based model to generate a synthetic dataset that transfers knowledge to the smaller on-device LM model.

\subsection{Textual Data Augmentation}
Textual data augmentation aims to efficiently expand the size of the training dataset by rephrasing or restructuring sentences, without the need for additional data collection and annotation.
 Early approaches typically involved word modification and syntactic manipulations in a sentence, such as synonym replacement, random insertion, random swapping, and random deletion \cite{belinkov2018synthetic-and-natural-noise, wei-zou-2019-eda}. 
More advanced methods leverage the text understanding capability of LLMs.
For instance, Back-translation \cite{sennrich-etal-2016-back-translation} proposed to use LM for translation, in which source sentences are translated into an intermediate language (e.g., German) and then translated back into the source language (e.g., English) to produce augmented samples in different words with similar underlying meanings.
While these traditional techniques have proven effective, they have the risk of generating redundant samples, thereby limiting improvements in dataset quality and diversity.

Recent progress in LLMs has shown superiority in producing high-quality, contextually relevant augmented text by analysing and reconstructing sentences to enhance both diversity and richness. 
For example, AugGPT \cite{Dai2025AugGPT} leverages the ChatGPT model to rewrite sentences while preserving dataset consistency, and Self-LLMDA \cite{li-etal-2024-empowering-self-LLMDA} automatically generates and selects optimal instructions to guide LLMs in producing augmented samples. 
Meanwhile, several studies \cite{Hongyuan2023EPA-easy-prompt-augment, Dawei2023DAIL-data-augment-self-paraphrase, chen-etal-2024-general-retrieval-llm-augment, Linmei2024llm-text-augment-personal-detections, HUanhuan2024text-classification-LLM-augmentation} have explored paraphrasing-based augmentation with LLMs, where the semantic content is preserved while surface forms, such as wording and structure, are altered. 
Such methods are particularly beneficial for scenarios with scarce data, as they need to enrich data without altering semantic meaning. 
However, these prompting-based techniques are not directly aligned with our objective of generating personalized text that reflects individual users’ preferences.

\section{Research Problem Formulation}
\begin{table*}
  \caption{List of Symbols used in this paper.}
  \label{tab:list_of_symbols}
  \resizebox{\textwidth}{!}{
  \begin{tabular}{ll}
    \toprule
    \textbf{Symbols} & \textbf{Description}\\
    \midrule
    \(x_u\)& The query of user \(u\). \\
    \(y_u\)& The response of user \(u\). \\
    \(t\)& The query time of a user. \\
    \(x^{syn}\)& The synthetic inputs. \\
    \(y^{syn}\)& The synthetic outputs. \\
    \(M_{base}\)& A task-specific pretrained model fine-tuned on public dataset without personal data. \\
    \(M_{device}\)& A personalized on-device language model fine-tuned from \(M_{base}\) with personal data. \\
    \(M_{cloud}\)& A pretrained cloud-based LLM with larger model size. \\
    \(M_{NLI}\)& A NLI model providing a semantic consistent score between synthetic and reference samples.\\
    $k$ & The number of samples generated by \(M_{cloud}\). \\
    \(D_u\) & The local dataset containing historical input-output pairs of user \(u\). \\
    \(D_u^{cls,syn}\)& The augmented dataset for classification tasks of user \(u\). \\
    \(D_u^{gen,syn}\)& The augmented dataset for generation tasks of user \(u\). \\
    \(D_u^{syn}\)& The augmented dataset included \(D_u^{cls,syn}\) and \(D_u^{gen,syn}\). \\
    \(D_u^{filtered}\)& The filtered high-quality augmented dataset of user \(u\). \\
    \(\epsilon_{scf} \)& The threshold to filter out dissimilar synthetic pairs. \\
    \(\epsilon_{tdf} \)& The threshold for the ROUGE-L score. \\
    \(\epsilon_{min\_len}\)& The threshold for minimum length. \\
    \(\epsilon_{max\_len}\)& The threshold for maximum length. \\
    $W_o^l$ & The pretrained parameters of $M_{base}$ at layer $l$. \\
    \(\Delta W_u^l\)& The trainable parameters in LoRA at layer $l$. \\
    $z^l$ & The input activation at layer $l$. \\
    $h^l$ & The output at layer $l$. \\
    $A^l, B^l$ & The LoRA rank-decomposition matrices at layer $l$. \\
    \(CE(\cdot)\)&  The cross-entropy loss function. \\
    \bottomrule
  \end{tabular}
  }
\end{table*}

This paper explores fine-tuning personalized on-device language models (LMs), incorporating two key concepts: personalized LMs and on-device LMs.
Unlike general LMs, which produce output sequences solely based on the input sequence, a personalized LM generates responses $y_{u,t}$ by considering both the user's query $x_{u,t}$ and their profile $D_{u}$.
We define the user profile as a collection of the user's historical input-output pairs:
i.e., $D_u=\{(x_{u,1}, y_{u,1}), (x_{u,2}, y_{u,2}), \dots, (x_{u,t-1}, y_{u,t-1})\}$, which represent the user’s interactions that occurred prior to the query time $t$.

Compared to a cloud-based LLM $M_{cloud}$, an on-device LM $M_{device}$ has a significantly smaller model size, as it must be deployed on a user's local device, where computational resources are limited. Additionally, unlike server-based LLMs, which are trained on extensive datasets collected from various sources, on-device LMs are only trained on a single user's data, which is often insufficient.
As a result, on-device LM has fast response ability, however, it does not have a powerful language understanding ability compared to server-based LLMs. In this paper, we aim to exploit both the on-device LM’s fast response merits and the strong inference ability of the server LLMs to build a high-performance on-device personalized LM.
For convenience, we summarize the key symbols and notations introduced in this paper in Table \ref{tab:list_of_symbols}.

\section{Proposed Method}
\begin{figure*}[t]
    \includegraphics[width=\textwidth]{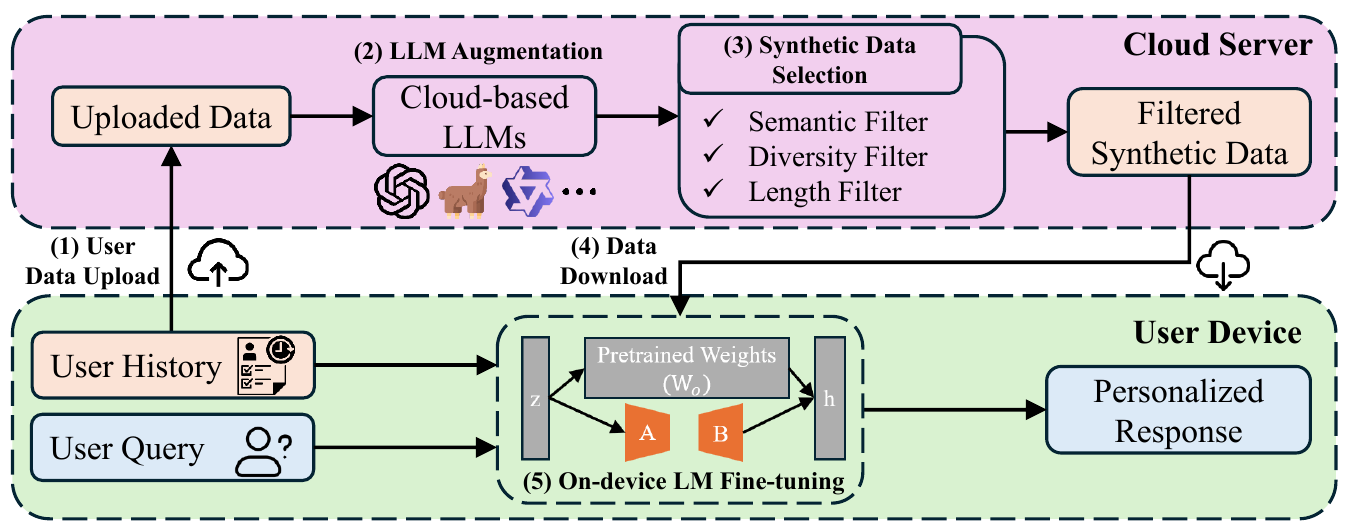}
    \Description{Overview of framework includes five steps. user data uploading, data augmentation with server LLM, synthetic data selection, synthetic data downloading, and on-device LM personalization fine-tuning.}
    \caption{Overview of the proposed method.}
    \label{fig:overview}
\end{figure*}

Different from previous work~\cite{salemi2023lamp}, which implements personalization in a cloud server setting, this paper proposes a cloud-device collaborative data augmentation for on-device personalized LM deployment framework that enhances inference efficiency without relying on the server LLM.
The basic idea of \modelname is to use the powerful server LLM model to assist the on-device personalized model's fine-tuning.
As shown in Figure \ref{fig:overview}, the proposed framework consists of the following five steps: (1) user data uploading, (2) data augmentation with server LLM, (3) synthetic data selection, (4) synthetic data downloading, and (5) on-device LM personalization fine-tuning. In the following parts, we provide a detailed description of each step.

\subsection{User Data Uploading} 
A user's historical profile provides a unique data distribution to capture users' preferences. 
However, due to the limited data size, directly fine-tuning the on-device model on this scarce local data cannot achieve satisfactory performance.
To address this issue, in \modelname, user upload their personal data to a central server, where a powerful cloud-based LLM performs data augmentation to generate additional personalized data.

\subsection{Data Augmentation with Server LLM}
On the server side, we prompt the cloud-based LLM to generate synthetic data tailored to the user’s preferences.
Since NLP tasks have different constraints on the output space, we categorize them into two major types, text classification and text generation, and design distinct augmentation strategies for each.

\textbf{Classification tasks.}  
In classification tasks, the output space is restricted to a predefined set of labels (e.g., movie tags, product ratings). 
Therefore, augmentation focuses on diversifying the input space while preserving the original output label. 
Specifically, given an input–output pair $(x_u, y_u)\in D_u$, the server LLM generates $k$ semantically similar inputs $\{x_{u,j}^{syn}\}_{j=1}^{k}$. 
The augmented dataset for classification tasks is defined as: 
\begin{equation}
    D_{u}^{cls,syn} = \{\{ (x_{u,j}^{syn},y_{u}) \}_{j=1}^{k} | (x_{u}, y_{u}) \in D_u\},
\end{equation}
where the augmented output remains unchanged, ensuring that augmented samples are in the task's predefined label space.

\textbf{Generation tasks.}  
In contrast, text generation tasks (e.g., news headline generation, title generation) lack a fixed label set. 
In this task, the server LLM not only generates synthetic inputs but also provides the corresponding outputs, effectively transferring its knowledge to the on-device model. 
Specifically, for each original pair $(x_u, y_u)\in D_u$, the server LLM generates $k$ semantically similar inputs $\{x_{u,j}^{syn}\}_{j=1}^{k}$ and their corresponding outputs $\{y_{u,j}^{syn}\}_{j=1}^{k}$. 
The augmented dataset for generation tasks is given by:
\begin{equation}
    D_{u}^{gen,syn} = \{\{ (x_{u,j}^{syn},y_{u,j}^{syn}) \}_{j=1}^{k} | (x_{u}, y_{u}) \in D_u\}.
\end{equation}
This process enables the small on-device LM to learn writing styles and knowledge from the larger teacher LLM.
Finally, the overall synthetic dataset can be expressed as:
\begin{equation}
    D_{u}^{syn} = 
    \begin{cases}
      D_{u}^{cls,syn}, & \text{if the task is classification,}\\
      D_{u}^{gen,syn}, & \text{if the task is generation.}\\
    \end{cases}    
\end{equation}

\begin{figure*}[t]
  \includegraphics[width=\textwidth]{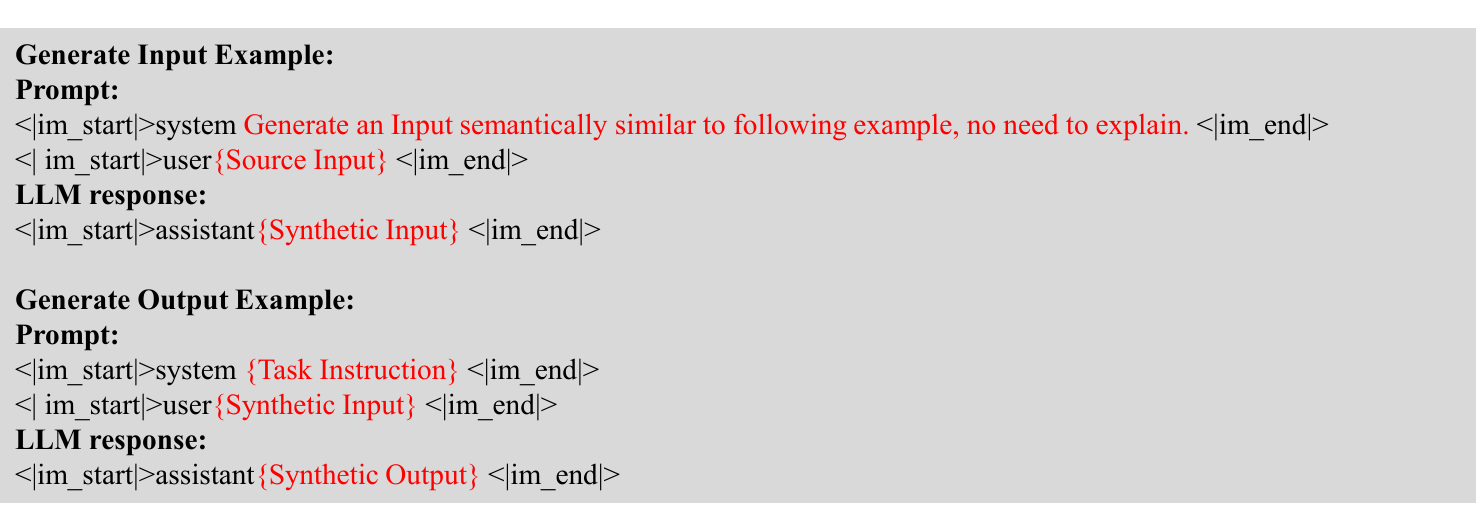}
  \Description{Using prompts to generate inputs, and passing synthetic inputs to generate outputs.}
  \caption{The illustration of the prompts used for personalized augmentation on server LLM. }
  \label{fig:augment prompt}
\end{figure*}

Figure \ref{fig:augment prompt} shows that we used the prompt and user query to synthesize new inputs, and then passed new inputs to produce responses.

\begin{algorithm}[t]
    % \SetAlgoNoLine
    \caption{Synthetic Data Selection}
    \label{alg:synthetic-selection}
    \DontPrintSemicolon
    \KwIn{
      User dataset $D_u = \{(x, y)\}$; synthetic dataset 
      $D^{\text{syn}}_u = \{\{(x^{\text{syn}}_j, y^{\text{syn}}_j)\}_{j=1}^k|(x,y)\}$; \\
      thresholds $\epsilon_{\text{scf}}$ (semantic), $\epsilon_{\text{tdf}}$ (diversity), 
      $\epsilon_{\min}$, $\epsilon_{\max}$ (length); \\
      $\mathsf{task} \in \{\text{classification},\text{generation}\}$; \\
    }
    \KwOut{
      Filtered synthetic dataset $D^{\text{filtered}}_u$
    }
    
    Initialize filtered dataset $D^{\text{filtered}}_u = \{\}$\;
    
    \For {\text{each} $(x, y) \in D_u$}{
    \For{$j=1$ \KwTo $k$}{
        $x^{\text{syn}} \leftarrow x^{\text{syn}}_{j}$; \quad $y^{\text{syn}} \leftarrow y^{\text{syn}}_{j}$\;
    
        \tcp{Filter 1: Semantic consistency (bidirectional NLI)}
        $\text{SCF} \leftarrow 
          \big[\mathrm{M_{NLI}}(x \Rightarrow x^{\text{syn}}) \ge \epsilon_{\text{scf}}\big]
          \wedge
          \big[\mathrm{M_{NLI}}(x^{\text{syn}} \Rightarrow x) \ge \epsilon_{\text{scf}}\big]$\;
    
        \tcp{Filter 2: Token diversity}
        $\text{TDF} \leftarrow \big[\mathrm{ROUGE\text{-}L}(x, x^{\text{syn}}) \le \epsilon_{\text{tdf}}\big]$\;
    
        \tcp{Filter 3: Length size}
        $r \leftarrow \dfrac{\text{len}(x^{\text{syn}})}{\text{len}(x)}$\;
        $\text{LSF} \leftarrow \big[\epsilon_{\min} \le r \le \epsilon_{\max}\big]$\;
    
        \If{$\text{SCF} \wedge \text{TDF} \wedge \text{LSF}$}{
          \eIf{$\mathsf{task} = \text{classification}$}{
            $D^{\text{filtered}}_u \leftarrow D^{\text{filtered}}_u \cup \{(x^{\text{syn}},\, y)\}$\;
          }{
            $D^{\text{filtered}}_u \leftarrow D^{\text{filtered}}_u \cup \{(x^{\text{syn}},\, y^{\text{syn}})\}$\;
          }
        }
    }
    }
    \Return $D^{\text{filtered}}_u$\;
\end{algorithm}

\subsection{Synthetic Data Selection}
Although $M_{cloud}$ generates a large amount of data for the target user, the generated data can be noisy, and not all samples contribute useful information for personalization fine-tuning.
Intuitively, high-quality augmented data should be similar to the original samples while still providing some diversity.
Therefore, we apply three carefully designed filters to select useful data for personalized LM.

\textbf{Filter 1: Semantic consistency filter.} 
Reliable synthetic data should preserve the semantics of the original statement without introducing hallucinated content. 
Natural Language Inference (NLI) models are trained to determine whether a ``hypothesis'' preserves the similar meaning as the ``premise''. 
Specifically, the NLI model classified the semantic relationship into one of three categories: (1) Entailment: The meaning of the hypothesis can be logically inferred from the premise. (2) Contradiction: The meaning of the hypothesis directly contradicts the premise. (3) Neutral: There is no clear logical connection or conflict between the premise and the hypothesis. 
For each given pairs $(x,x_{syn})$, semantic evaluator outputs an entailment probability in both direction, $M_{NLI}(x \Rightarrow x_{syn})$ and $M_{NLI}( x_{syn} \Rightarrow x)$. 
We treat the sample as semantically consistent if both probabilities exceed a threshold $\epsilon_{scf}$.
Therefore, we employ a small NLI model $M_{NLI}$ \citep{liu-etal-2022-wanli} as the semantic evaluator and define the filters as:
\begin{equation}
      SCF =(M_{NLI}(x \Rightarrow x_{syn}) \geq \epsilon_{scf}) \wedge \\
          (M_{NLI}( x_{syn} \Rightarrow x) \geq \epsilon_{scf}),
  \end{equation}
where $\epsilon_{scf}$ is the threshold to filter out dissimilar synthetic pairs, and $M_{NLI}(a \Rightarrow b)$ indicates the possibility of inferring $b$ given $a$.

\textbf{Filter 2: Token diversity filter.}  
While the SCF filter ensures the consistency of semantics for synthetic data, it is also important to maintain diversity in the augmented data. 
Ideally, synthetic samples should convey the original meaning but with different wording.
To measure this, we apply the ROUGE-L \citep{lin-2004-rouge} metric to assess token overlap between original and generated sequences:
\begin{equation}
  TDF = \operatorname{ROUGE-L}(x, x_{syn}) \leq \epsilon_{tdf},
\end{equation}
where $\epsilon_{tdf}$ is the threshold for the ROUGE-L score.

\textbf{Filter 3: Length size filter.}
Finally, we ensure that synthetic samples have a reasonable length to avoid abnormal or redundant data. We discard data that are either too short or too long, using predefined minimum and maximum length thresholds $\epsilon_{min\_len}$ and $\epsilon_{max\_len}$:
\begin{equation}
  LSF=(len(x_{syn}) \geq \epsilon_{min\_len}\cdot len(x) \wedge \\
      (len(x_{syn}) \leq \epsilon_{max\_len}\cdot len(x)),
\end{equation}
Specifically, we filter all generated samples whose length ratio (i.e., the length ratio of $x_{syn}$ to $x$) is out of the pre-defined range $[\epsilon_{min\_len}, \epsilon_{max\_len}]$ to ensure the generated sample has a length similar to the input.
By applying these three filters, we obtain a high-quality dataset $D_{filtered}$ from the synthetic data pool $D_{syn}$, which is then used for LM fine-tuning. 
Algorithm \ref{alg:synthetic-selection} illustrates the detailed steps of synthetic data selection.

\subsection{On-device LM Deployment}
\subsubsection{\textbf{Synthetic Data Downloading}}
After selecting the high-quality augmented data $D_{filtered}$, the server sends these data back to the corresponding users. Users then download the data and combine it with their local datasets for on-device fine-tuning.

\subsubsection{\textbf{On-device LM Personalization Fine-tuning}}
We employ a pretraining and efficient fine-tuning approach for on-device personalization.
Specifically, for a target task, we fine-tune a general LM on a public, standard dataset to enhance its general task understanding. 
Since this step does not involve personal data, it is executed on the cloud server to avoid using the constrained on-device resources. 
After optimization, we obtain a task-specific pretrained model $M_{base}$ as the initialization point for personalized LMs fine-tuning.

On the device, we fine-tune an on-device LM $M_{base}$ on the combined synthetic and user local datasets to learn both personalized information from the user and the insightful knowledge from the server LLM. The on-device LM is much smaller than the server LLM, ensuring low inference latency.

To reduce training costs, we implement parameter-efficient fine-tuning (PEFT) using LoRA \citep{dettmers2023qloraefficientfinetuningquantized}.
It introduces trainable adapters $\Delta W_u$ into the pretrained parameters $W_o$ of $M_{base}$. Since $W_o$ contains significantly more parameters than $\Delta W_u$, we freeze $W_o$ and only fine-tune on-device LM $M_{base}$ by optimizing $\Delta W_u$ for personalization.
LoRA decomposes the pretrained weights $W_o\in \mathbb{R} ^{d\times k}$  into two smaller weight matrices $A\in \mathbb{R} ^{d\times r}$ and $B\in \mathbb{R} ^{r\times k}$, where r is the rank of LoRA matrix and is much smaller than $d$ and $k$. 
For the output of $l$-th linear layer, $h^l=W_o^lz^l$, LoRA modifies it to:
\begin{equation}
    h^l=W_o^lz^l+\Delta W^l_uz^l=W_o^lz^l+B^lA^lz^l,
\end{equation}
where $z^l\in \mathbb{R} ^{k}$ denotes the input activation layer at layer $l$.
Therefore, it is efficient to train a small number of parameters in LoRA matrices.
At the beginning, $A$ is initialized with random Gaussian and $B$ is initialized with zero.
Finally, the trainable adapters $\Delta W_u$ integrates into base LM $M_{base}$, forming the personalized LM $M_{device}$:
\begin{equation}
    M_{device} = M_{base} + \Delta W_u,
\end{equation}
We then only optimize $\Delta W_u$ using the user's historical data $D_{u}$ and the filtered LLM-generated data $D_{u}^{filtered}$. 
\begin{equation}
    \Delta W_u = \mathop{argmin}CE(M_{device} |D_{u} \cup D_{u}^{filtered}),
  \end{equation}
where $CE(\cdot)$ represents the cross-entropy loss function.

After optimizing the personalized LM $M_{device}$, users can then process queries locally without relying on the cloud server, benefiting from lower latency and without relying on network connection.

\section{Experiments}
In this section, we present the experimental setup and the results, followed by an in-depth analysis.
we conduct extensive experiments aiming to address the following research questions (RQs):
\begin{itemize}
    \item \textbf{RQ1:} How does \modelname perform compared with existing personalization methods?
    \item \textbf{RQ2:} What's the impact of the specifically designed personalized data augmentation and selection strategies on the performance of \modelname?
    \item \textbf{RQ3:} How does the size of augmentation data influence the performance of \modelname?
    \item \textbf{RQ4:} How's the computational efficiency of \modelname?
    \item \textbf{RQ5:} How's the generalization ability of our framework with variant LLMs?
    \item \textbf{RQ6:} Does the generated synthetic data align with users' intent and preferences?
\end{itemize}

% \caption{The statistics of datasets used in our experiment. It presents the total number of user query (#Q) and history (#History). $L_{in}$ and $L_{out}$ are the average tokens of inputs and outputs.}

% \usepackage{multirow}
% \usepackage{graphicx}
\begin{table}[t]
\caption{Statistics of the preprocessed datasets..}
\label{tab:dataset}
\centering
\begin{tabular}{lccccccc}
\hline
\multicolumn{1}{c}{\multirow{2}{*}{\textbf{Task}}} & \multicolumn{4}{c|}{\textbf{Target Users}} & \multicolumn{3}{c}{\textbf{Filtered Dataset}} \\
\multicolumn{1}{c}{} & \multicolumn{1}{l}{\# Q} & \multicolumn{1}{l}{\# History} & \multicolumn{1}{l}{$L_{in}$} & \multicolumn{1}{l|}{$L_{out}$} & \multicolumn{1}{l}{\# Q} & \multicolumn{1}{l}{$L_{in}$} & \multicolumn{1}{l}{$L_{out}$} \\ \hline
\textbf{LaMP-1} & 100 & 317.57 & 78.43 & 3.0 & 15928 & 161.76 & 19.15 \\
\textbf{LaMP-2} & 2752 & 54.58 & 129.55 & 2.24 & 8962 & 121.21 & 2.20 \\
\textbf{LaMP-3} & 100 & 959.02 & 244.79 & 1.00 & 15721 & 193.19 & 1.00 \\
\textbf{LaMP-4} & 955 & 269.08 & 31.49 & 15.60 & 12736 & 27.40 & 14.27 \\
\textbf{LaMP-5} & 100 & 443.03 & 222.60 & 15.52 & 23473 & 156.92 & 18.63 \\
\textbf{LaMP-7} & 100 & 120.15 & 40.90 & 27.66 & 16490 & 39.56 & 0.00 \\ \hline
\end{tabular}
\end{table}
\subsection{Experimental Settings}
\subsubsection{\textbf{Datasets.}}
To validate the effectiveness of the proposed method, we conduct extensive experiments on six personalization tasks in Large Language Model Personalization (LaMP) benchmark \citep{salemi2023lamp}, including three classification tasks (LaMP-1: Personalized Citation Identification, LaMP-2: Personalized Movie Tagging, LaMP-3: Personalized Product Rating) and three generation tasks (LaMP-4: Personalized News Headline Generation, LaMP-5: Personalized Scholarly Title Generation, and LaMP-7: Personalized Tweet Paraphrasing). \footnote{We exclude the LaMP-6: Email Subject Generation task as it relies on private data that we cannot access.}
In this study, we use the time-based separation data in the LaMP benchmark. 
Note that our data usage slightly differs from the original LaMP benchmark setting. Specifically, they address multi-user personalization by training models on the combined histories of multiple users, which results in coarse-grained personalization. 
In contrast, our work focuses on fine-grained personalization by training a separate model for each individual user using only their own historical data.
The core statistics of processed datasets for each task are presented in Table \ref{tab:dataset}. \#Q and \#History represent the total number of user queries and history, respectively, in the target users' test dataset and the synthetic selected training dataset. $L_{in}$ and $L_{out}$ are the average tokens of inputs and outputs.

To promote the personalization phenomenon, we designate two groups of target users: the 100 most active users with the longest history logs and an additional 100 users randomly sampled from the dataset, while using all remaining users for base model training. 
Our objective is to obtain an on-device personalized LM for each user among these 200 users. 
Sampling strategies are widely adopted to reduce the computational overhead of personalization evaluation, as demonstrated in many existing works evaluated on the LaMP benchmark.
For instance, HYDRA \cite{Zhuang2024HYDRA} randomly selected 100 users for training and an additional 50 users for evaluation, while FERMI \cite{kim2025fewshotpersonalization} selected 50 users for training and evaluation. 
Similarly, OPPU \cite{tan2024OPPUPEFTperuser}, Personaized Piece \cite{tan2024personalizedpiecesPEFT}, and PriME \cite{kim2025personalizedLMmerging} selected the 100 most active users with the longest history logs for training and evaluation. 
In real-world deployment, these on-device personalized LMs can be executed in parallel and independently on each user's device. However, for our evaluation experiments, we have to build these user-specific personalized models one by one due to the equipment limitation, leading to a relatively long evaluation time. Therefore, we have to sample only a part of the users for evaluation. We believe this may also be the main reason that previous works use sampling user strategies rather than full-set users for evaluation.
In this paper, we not only follow the same sampling strategies of prior works \cite{tan2024OPPUPEFTperuser,tan2024personalizedpiecesPEFT,kim2025personalizedLMmerging}, but also adopt a random sampling approach, which offers a statistically representative assessment of overall dataset performance.

\subsubsection{\textbf{Evaluation Metrics.}}
Following LaMP \citep{salemi2023lamp}, we use accuracy and F1-score for the LaMP-1 and LaMP-2, mean absolute error (MAE) and root mean square error (RMSE) for the LaMP-3, and ROUGE-1, ROUGE-L, and BERTScore-F1 (BERT-F1) for LaMP-4, LaMP-5, and LaMP-7. 
Except for MAE and RMSE, where lower values are better, all other metrics with higher values indicate better performance.
The mathematical equations of metrics are presented in Appendix \ref{metric_equation}

ROUGE-N \citep{lin-2004-rouge}, or Recall-Oriented Understudy for Gisting Evaluation, measures the overlap of N-grams, sequences of N words, between the prediction and reference texts. 
ROUGE-1 measures the overlap of each word (unigrams) between the prediction and reference. 
ROUGE-L is based on the longest common subsequence (LCS) between prediction and reference, which considers sentence-level structure similarity and identifies the longest co-occurring words in texts.
BERTScore \citep{bertscore_zhang2020} computes a similarity score for each token in the prediction with each token in the reference text. 
It leverages the pre-trained contextual embeddings from BERT models and matches words in prediction and reference texts by cosine similarity. 
In this paper, we report the F1-score from BERTScore.

\subsubsection{\textbf{Baselines.}}
We compare \modelname with the non-personalized models and other personalized baselines. In the non-personalized baselines, we select the cloud-based LLM ($M_{cloud}$) and on-device LM ($M_{device}$), which are fine-tuned only on the remaining users without the $100$ target users.

The personalized baselines include RAG-based methods, fine-tuning-based methods, and the speculative decoding method. For fair comparison, these personalized methods are all implemented on on-device models:
\begin{itemize}
    \item Retrieval-Augmented Personalization (RAG): RAG incorporates relevant items from the target user's history to the prompt \cite{salemi2023lamp} to achieve a personalized response. To showcase the deteriorated performance of RAG in on-device LM, we also present the performance of RAG in the cloud counterpart. Following \cite{salemi2023lamp}, in the experiment, we implement a sparse retriever: BM25 \citep{BM25retrieval2014} and a dense retriever: Contriever \citep{izacard2021contriever}.
    \item Speculative Decoding: An optimization technique accelerates models' inference by using a non-personalized LLM on the cloud to assist a personalized LM on the device. We evaluate personalized tasks on Google's speculative decoding \citep{Speculative_decode_2023} implemented by HuggingFace \citep{gante2023hugging_speculative}.
    \item Direct-FT \cite{tan2024OPPUPEFTperuser}: Directly LoRA fine-tuning $M_{device}$ uses the target user's local historical data. This method cannot be satisfied due to the limited local data size.
    \item EDA-FT \cite{wei-zou-2019-eda}: EDA (Easy Data Augmentation) is a traditional text data augmentation method including synonym replacement, random insertion, random swap, and random deletion.
    \item RKD-FT: An LLM knowledge distillation method uses reverse KL divergence
\cite{gu2024minillmknowledgedistillationlarge}.
\end{itemize}
For EDA-FT, RKD-FT, and \modelname, they augment local knowledge based on users' data.

\subsubsection{\textbf{Implementation.}}
For all baselines in our study, we choose models from one of the most widely adopted open-source LLM series \textit{Qwen2.5} \footnote{\url{https://github.com/QwenLM/Qwen2.5}} \citep{qwen2.5}. 
Specifically, we use \textit{Qwen2.5-3B-Instruct} as the cloud-based model and \textit{Qwen2.5-0.5B-Instruct} as the on-device model for each user. 
To ensure efficiency, we choose one retriever item for all retrieval-based methods.

By default, we set the LLM generation samples $k$ to 5 in all experiments. 
For each case, we independently execute the model five times with different random seeds
and report the averaged outcomes, and all results are statistically significant at $p <$ 0.05.
We apply the LoRA adapter on all linear layers of the on-device model, and set the LoRA rank $r$ to 16 and scaling factor $\alpha$ to 8.
We quantize the on-device model weight in the NF4 data type and use bfloat16 for computation.
Followed by the Qwen2.5 technique report \citep{qwen2.5}, we used the multinomial sampling decoding with temperature $\tau_{temp}=0.7$ to balance the computational efficiency and sampling diversity of data generation. 
We implement all the experiments using Pytorch \cite{paszke2019pytorch-library} and HuggingFace library \cite{wolf2020huggingfaces} on an NVIDIA RTX A5000 GPU.

% main experiment 
% \caption{The performance of \modelname and baseline on LaMP benchmark. The best performance of personalized on-device model $M_{device}$ is highlighted in \textbf{bold} and the second best is \underline{underlined}.}
% \label{tab:main_exp}
% $M_{cloud}$ $M_{device}$
% \uparrow \downarrow

% \usepackage{multirow}
% \usepackage{graphicx}
% \usepackage[normalem]{ulem}
% \useunder{\uline}{}{ul} removed underlined as not approved package 
\begin{table*}[t]
\caption{The performance of \modelname and baselines on LaMP benchmark for the \textbf{top 100 most active users}. The best performance of personalized on-device model $M_{device}$ is highlighted in \textbf{bold}.}
\label{tab:main_exp}
\resizebox{\textwidth}{!}{%
\begin{tabular}{llcclcccclclclclcc}
\hline
\multicolumn{1}{c}{\multirow{2}{*}{\textbf{Tasks}}} & \multicolumn{1}{c}{\multirow{2}{*}{\textbf{Metric}}} & \multicolumn{2}{c}{\textbf{Non-Personalized}} &  & \multicolumn{4}{c}{\textbf{RAG}} &  & \textbf{\begin{tabular}[c]{@{}c@{}}Speculative \\ Decoding\end{tabular}} &  & \multicolumn{6}{c}{\textbf{Fine-tuning based  $M_{device}$}} \\ \cline{3-4} \cline{6-9} \cline{11-11} \cline{13-18} 
\multicolumn{1}{c}{} & \multicolumn{1}{c}{} & $M_{cloud}$ & $M_{device}$ & \multicolumn{1}{l}{} & \begin{tabular}[c]{@{}c@{}}$M_{cloud}$\\ +BM25\end{tabular} & \begin{tabular}[c]{@{}c@{}}$M_{device}$\\ +BM25\end{tabular} & \begin{tabular}[c]{@{}c@{}}$M_{cloud}$\\ +Contriever\end{tabular} & \begin{tabular}[c]{@{}c@{}}$M_{device}$\\ +Contriever\end{tabular} & \multicolumn{1}{l}{} & $M_{device}$ & \multicolumn{1}{l}{} & \begin{tabular}[c]{@{}c@{}}Direct\\ -FT\end{tabular} & \multicolumn{1}{l}{} & \begin{tabular}[c]{@{}c@{}}EDA\\ -FT\end{tabular} & \multicolumn{1}{l}{} & \begin{tabular}[c]{@{}c@{}}RKD\\ -FT\end{tabular} & \begin{tabular}[c]{@{}c@{}}CDCDA\\ -PLM\end{tabular} \\ \hline
\multirow{2}{*}{\textbf{LaMP-1}} & Accuracy $\uparrow$ & 0.520 & 0.390 & \multicolumn{1}{l|}{} & 0.560 & 0.310 & 0.630 & 0.230 &  & { 0.500} & \multicolumn{1}{l|}{} & 0.420 &  & 0.410 &  & \multicolumn{1}{c|}{0.460} & \textbf{0.530} \\
 & F1 $\uparrow$ & 0.515 & 0.356 & \multicolumn{1}{l|}{} & 0.528 & 0.381 & 0.605 & 0.245 &  & { 0.469} & \multicolumn{1}{l|}{} & 0.390 &  & 0.382 &  & \multicolumn{1}{c|}{0.389} & \textbf{0.483} \\ \hline
\multirow{2}{*}{\textbf{LaMP-2}} & Accuracy $\uparrow$ & 0.248 & 0.017 & \multicolumn{1}{l|}{} & 0.319 & 0.009 & 0.292 & 0.050 &  & { 0.353} & \multicolumn{1}{l|}{} & 0.243 &  & 0.296 &  & \multicolumn{1}{c|}{0.283} & \textbf{0.391} \\
 & F1 $\uparrow$ & 0.129 & 0.017 & \multicolumn{1}{l|}{} & 0.225 & 0.019 & 0.234 & 0.066 &  & { 0.201} & \multicolumn{1}{l|}{} & 0.099 &  & 0.156 &  & \multicolumn{1}{c|}{0.125} & \textbf{0.224} \\ \hline
\multirow{2}{*}{\textbf{LaMP-3}} & MAE $\downarrow$ & 1.120 & 0.640 & \multicolumn{1}{l|}{} & 1.970 & 1.580 & 1.630 & 1.465 &  & 0.550 & \multicolumn{1}{l|}{} & 0.474 &  & { 0.450} &  & \multicolumn{1}{c|}{0.474} & \textbf{0.400} \\
 & RMSE $\downarrow$ & 1.371 & 1.131 & \multicolumn{1}{l|}{} & 2.508 & 2.191 & 2.252 & 2.117 &  & 1.034 & \multicolumn{1}{l|}{} & 0.946 &  & \textbf{0.831} &  & \multicolumn{1}{c|}{0.912} & { 0.834} \\ \hline
\multirow{3}{*}{\textbf{LaMP-4}} & ROUGE-1 $\uparrow$ & 0.107 & 0.102 & \multicolumn{1}{l|}{} & 0.122 & 0.092 & 0.121 & 0.098 &  & 0.110 & \multicolumn{1}{l|}{} & 0.106 &  & { 0.117} &  & \multicolumn{1}{c|}{0.116} & \textbf{0.120} \\
 & ROUGE-L $\uparrow$ & 0.096 & 0.090 & \multicolumn{1}{l|}{} & 0.110 & 0.083 & 0.109 & 0.088 &  & 0.098 & \multicolumn{1}{l|}{} & 0.094 &  & { 0.104} &  & \multicolumn{1}{c|}{0.103} & \textbf{0.107} \\
 & BERT-F1 $\uparrow$ & 0.847 & 0.838 & \multicolumn{1}{l|}{} & 0.849 & 0.837 & 0.850 & 0.839 &  & { 0.847} & \multicolumn{1}{l|}{} & 0.845 &  & { 0.847} &  & \multicolumn{1}{c|}{{ 0.847}} & \textbf{0.849} \\ \hline
\multirow{3}{*}{\textbf{LaMP-5}} & ROUGE-1 $\uparrow$ & 0.427 & 0.360 & \multicolumn{1}{l|}{} & 0.457 & 0.328 & 0.453 & 0.346 &  & 0.341 & \multicolumn{1}{l|}{} & { 0.375} &  & 0.370 &  & \multicolumn{1}{c|}{{ 0.375}} & \textbf{0.382} \\
 & ROUGE-L $\uparrow$ & 0.362 & 0.309 & \multicolumn{1}{l|}{} & 0.379 & 0.292 & 0.387 & 0.292 &  & 0.279 & \multicolumn{1}{l|}{} & 0.314 &  & { 0.316} &  & \multicolumn{1}{c|}{0.307} & \textbf{0.317} \\
 & BERT-F1 $\uparrow$ & 0.894 & 0.885 & \multicolumn{1}{l|}{} & 0.896 & 0.882 & 0.896 & 0.878 &  & 0.879 & \multicolumn{1}{l|}{} & \textbf{0.886} &  & { 0.885} &  & \multicolumn{1}{c|}{0.884} & \textbf{0.886} \\ \hline
\multirow{3}{*}{\textbf{LaMP-7}} & ROUGE-1 $\uparrow$ & 0.365 & 0.337 & \multicolumn{1}{l|}{} & 0.355 & 0.296 & 0.338 & 0.230 &  & 0.354 & \multicolumn{1}{l|}{} & 0.337 &  & 0.373 &  & \multicolumn{1}{c|}{{ 0.374}} & \textbf{0.383} \\
 & ROUGE-L $\uparrow$ & 0.310 & 0.297 & \multicolumn{1}{l|}{} & 0.315 & 0.262 & 0.296 & 0.201 &  & 0.317 & \multicolumn{1}{l|}{} & 0.302 &  & 0.327 &  & \multicolumn{1}{c|}{{ 0.328}} & \textbf{0.336} \\
 & BERT-F1 $\uparrow$ & 0.881 & 0.877 & \multicolumn{1}{l|}{} & 0.881 & 0.869 & 0.879 & 0.854 &  & 0.878 & \multicolumn{1}{l|}{} & 0.875 &  & 0.880 &  & \multicolumn{1}{c|}{\textbf{0.882}} & { 0.881} \\ \hline
\end{tabular}
}
\end{table*}
\begin{table*}[t]
\caption{The performance of \modelname and baselines on LaMP benchmark for \textbf{100 randomly sampled users}. The best performance of personalized on-device model $M_{device}$ is highlighted in \textbf{bold}.}
\label{tab:ran_users_exp}
\resizebox{\textwidth}{!}{%
\begin{tabular}{llcccccccccccccccc}
\hline
\multicolumn{1}{c}{\multirow{2}{*}{\textbf{Tasks}}} & \multicolumn{1}{c}{\multirow{2}{*}{\textbf{Metric}}} & \multicolumn{2}{c}{\textbf{Non-Personalized}} & \multicolumn{1}{l}{} & \multicolumn{4}{c}{\textbf{RAG}} & \multicolumn{1}{l}{} & \textbf{\begin{tabular}[c]{@{}c@{}}Speculative \\ Decoding\end{tabular}} & \multicolumn{1}{l}{} & \multicolumn{6}{c}{\textbf{Fine-tuning based  $M_{device}$}} \\ \cline{3-4} \cline{6-9} \cline{11-11} \cline{13-18} 
\multicolumn{1}{c}{} & \multicolumn{1}{c}{} & $M_{cloud}$ & $M_{device}$ & \multicolumn{1}{l}{} & \begin{tabular}[c]{@{}c@{}}$M_{cloud}$\\ +BM25\end{tabular} & \begin{tabular}[c]{@{}c@{}}$M_{device}$\\ +BM25\end{tabular} & \begin{tabular}[c]{@{}c@{}}$M_{cloud}$\\ +Contriever\end{tabular} & \begin{tabular}[c]{@{}c@{}}$M_{device}$\\ +Contriever\end{tabular} & \multicolumn{1}{l}{} & $M_{device}$ & \multicolumn{1}{l}{} & \begin{tabular}[c]{@{}c@{}}Direct\\ -FT\end{tabular} & \multicolumn{1}{l}{} & \begin{tabular}[c]{@{}c@{}}EDA\\ -FT\end{tabular} & \multicolumn{1}{l}{} & \begin{tabular}[c]{@{}c@{}}RKD\\ -FT\end{tabular} & \begin{tabular}[c]{@{}c@{}}CDCDA\\ -PLM\end{tabular} \\ \hline
\multirow{2}{*}{\textbf{LaMP-1}} & Accuracy $\uparrow$ & 0.480 & 0.500 & \multicolumn{1}{c|}{} & 0.530 & 0.480 & 0.550 & 0.500 &  & 0.480 & \multicolumn{1}{c|}{} & { 0.520} &  & { 0.520} &  & \multicolumn{1}{c|}{0.470} & \textbf{0.580} \\
 & F1 $\uparrow$ & 0.466 & 0.469 & \multicolumn{1}{c|}{} & 0.487 & 0.324 & 0.502 & 0.333 &  & 0.442 & \multicolumn{1}{c|}{} & 0.473 &  & { 0.513} &  & \multicolumn{1}{c|}{0.435} & \textbf{0.566} \\ \hline
\multirow{2}{*}{\textbf{LaMP-2}} & Accuracy $\uparrow$ & 0.155 & 0.010 & \multicolumn{1}{c|}{} & 0.224 & 0.035 & 0.255 & 0.052 &  & { 0.169} & \multicolumn{1}{c|}{} & 0.072 &  & 0.103 &  & \multicolumn{1}{c|}{0.110} & \textbf{0.217} \\
 & F1 $\uparrow$ & 0.139 & 0.013 & \multicolumn{1}{c|}{} & 0.233 & 0.044 & 0.258 & 0.081 &  & { 0.114} & \multicolumn{1}{c|}{} & 0.065 &  & 0.076 &  & \multicolumn{1}{c|}{0.070} & \textbf{0.169} \\ \hline
\multirow{2}{*}{\textbf{LaMP-3}} & MAE $\downarrow$ & 1.100 & 0.790 & \multicolumn{1}{c|}{} & 1.540 & 1.410 & 0.950 & 1.010 &  & 0.580 & \multicolumn{1}{c|}{} & 0.580 &  & { 0.480} &  & \multicolumn{1}{c|}{0.610} & \textbf{0.460} \\
 & RMSE $\downarrow$ & 1.386 & 1.330 & \multicolumn{1}{c|}{} & 2.159 & 1.972 & 1.584 & 1.559 &  & 1.149 & \multicolumn{1}{c|}{} & 1.105 &  & \textbf{0.980} &  & \multicolumn{1}{c|}{1.171} & { 1.000} \\ \hline
\multirow{3}{*}{\textbf{LaMP-4}} & ROUGE-1 $\uparrow$ & 0.133 & 0.128 & \multicolumn{1}{c|}{} & 0.151 & 0.122 & 0.159 & 0.122 &  & 0.145 & \multicolumn{1}{c|}{} & 0.136 &  & 0.111 &  & \multicolumn{1}{c|}{\textbf{0.149}} & { 0.148} \\
 & ROUGE-L $\uparrow$ & 0.117 & 0.113 & \multicolumn{1}{c|}{} & 0.134 & 0.109 & 0.143 & 0.110 &  & 0.125 & \multicolumn{1}{c|}{} & 0.124 &  & 0.099 &  & \multicolumn{1}{c|}{{ 0.131}} & \textbf{0.133} \\
 & BERT-F1 $\uparrow$ & 0.843 & 0.841 & \multicolumn{1}{c|}{} & 0.848 & 0.842 & 0.850 & 0.841 &  & 0.844 & \multicolumn{1}{c|}{} & 0.844 &  & 0.833 &  & \multicolumn{1}{c|}{{ 0.846}} & \textbf{0.847} \\ \hline
\multirow{3}{*}{\textbf{LaMP-5}} & ROUGE-1 $\uparrow$ & 0.438 & 0.365 & \multicolumn{1}{c|}{} & 0.448 & 0.356 & 0.464 & 0.343 &  & 0.354 & \multicolumn{1}{c|}{} & 0.363 &  & 0.343 &  & \multicolumn{1}{c|}{{ 0.371}} & \textbf{0.376} \\
 & ROUGE-L $\uparrow$ & 0.370 & 0.301 & \multicolumn{1}{c|}{} & 0.375 & 0.286 & 0.386 & 0.289 &  & 0.313 & \multicolumn{1}{c|}{} & 0.311 &  & 0.288 &  & \multicolumn{1}{c|}{{ 0.318}} & \textbf{0.324} \\
 & BERT-F1 $\uparrow$ & 0.892 & { 0.885} & \multicolumn{1}{c|}{} & 0.896 & 0.883 & 0.897 & 0.881 &  & { 0.885} & \multicolumn{1}{c|}{} & 0.880 &  & 0.884 &  & \multicolumn{1}{c|}{0.883} & \textbf{0.889} \\ \hline
\multirow{3}{*}{\textbf{LaMP-7}} & ROUGE-1 $\uparrow$ & 0.352 & 0.257 & \multicolumn{1}{c|}{} & 0.323 & 0.214 & 0.326 & 0.225 &  & 0.330 & \multicolumn{1}{c|}{} & 0.294 &  & 0.321 &  & \multicolumn{1}{c|}{{ 0.332}} & \textbf{0.354} \\
 & ROUGE-L $\uparrow$ & 0.295 & 0.213 & \multicolumn{1}{c|}{} & 0.271 & 0.182 & 0.276 & 0.196 &  & 0.269 & \multicolumn{1}{c|}{} & 0.241 &  & 0.272 &  & \multicolumn{1}{c|}{{ 0.277}} & \textbf{0.295} \\
 & BERT-F1 $\uparrow$ & 0.884 & 0.862 & \multicolumn{1}{c|}{} & 0.882 & 0.856 & 0.884 & 0.858 &  & 0.881 & \multicolumn{1}{c|}{} & 0.877 &  & 0.881 &  & \multicolumn{1}{c|}{{ 0.882}} & \textbf{0.884} \\ \hline
\end{tabular}%
}
\end{table*}
\subsection{Overall Results (RQ1)}
To validate our proposed method's effectiveness, we compare it with several baselines and show the results in Table \ref{tab:main_exp}. From the results, we have some interesting observations as follows.

First, by comparing the cloud model $M_{cloud}$ with the device model $M_{device}$, we observe that the cloud model performs significantly better than the corresponding device model in both personalized and non-personalized settings. This is because cloud-based models have a much larger number of parameters, approximately six times more in our experiments, making them unsuitable for deployment on edge devices. This finding highlights the necessity of transferring knowledge from the cloud-based LLM to support the weaker on-device LM.

Furthermore, when comparing RAG-based personalization methods, we find that the performance of the small on-device model actually declines after incorporating RAG. This aligns with our argument that on-device models are too small to effectively support prompt-based personalization.

By comparing speculative decoding with other baselines, we observe that it achieves relatively strong performance. However, as discussed in the baseline section, speculative decoding relies on frequent interaction with a cloud-based LLM, which introduces additional latency and requires a stable network connection. 
In addition, our proposed \modelname outperforms speculative decoding by a clear margin, demonstrating its superiority in terms of performance. 
Moreover, \modelname operates entirely on-device without the need for network connectivity and enables fast decoding, offering further advantages over the speculative decoding approach.

Among fine-tuning-based personalization approaches, Direct-FT yields the worst performance due to the limited availability of local user data, which is typically insufficient for effective personalized fine-tuning. The baseline methods, EDA-FT and RKD-FT, improve upon direct fine-tuning in some tasks, but their enhancements are limited. In some cases, their performance even deteriorates, likely due to the simplistic knowledge augmentation techniques they employ.

In Table \ref{tab:main_exp}, we selected the top 100 most active users followed by OPPU \citep{tan2024OPPUPEFTperuser} and Personalized Piece \citep{tan2024personalizedpiecesPEFT}. 
To better validate the effectiveness of our method, due to the limited computational resources, we conducted an additional experiment on all baselines using 100 randomly selected users from each task.
Since these users were chosen at random, the results are expected to be representative of a broader user base. 
The experimental results, Table \ref{tab:ran_users_exp}, demonstrate that our method outperforms nearly all baselines across all tasks on additional users.

Our proposed \modelname consistently outperforms all on-device baselines across all tasks. Additionally, \modelname achieves performance comparable to cloud models, demonstrating its effectiveness and strong generalization ability.

% ablation
% \caption{Ablation studies results with respect to server LLM data augmentation (LDA) and data selection (DS) components. The best performances are highlighted in \textbf{bold}.}

% \usepackage{multirow}
% \usepackage{graphicx}
\begin{table}[t]
\caption{Ablation studies results with respect to server LLM data augmentation (LDA) and data selection (DS) components. The best are highlighted in \textbf{bold}.}
\label{tab:ablation_components}
\centering
\begin{tabular}{lcc|cc}
\hline
\multicolumn{1}{c}{\multirow{2}{*}{\textbf{Methods}}} & \multicolumn{2}{c|}{\textbf{LaMP-2}} & \multicolumn{2}{c}{\textbf{LaMP-7}} \\
\multicolumn{1}{c}{} & Acc & F1 & \multicolumn{1}{l}{R-1} & \multicolumn{1}{l}{R-L} \\ \hline
Full model (DS) & \textbf{0.391} & \textbf{0.224} & \textbf{0.383} & \textbf{0.336} \\ \hline
Full model (RS) & 0.324 & 0.157 & 0.344 & 0.302 \\
-DS & 0.360 & 0.187 & 0.354 & 0.314 \\ \hline
-DS -LDA & 0.243 & 0.099 & 0.337 & 0.302 \\ \hline
-DS -LDA -FT & 0.017 & 0.017 & 0.337 & 0.297 \\ \hline
\end{tabular}%
\end{table}
\subsection{Ablation Study (RQ2)}
In this part, on LaMP-2 and LaMP-7, we demonstrate the effectiveness of our delicately designed modules in \modelname, including LLM data augmentation (LDA) and data selection (DS) components. 

As shown in Table \ref{tab:ablation_components}, when we replace our carefully designed filters in DS with random selection (RS), the accuracy of the full model with DS drops from 0.391 to 0.324 on LaMP-2.

When we remove the DS (i.e., -DS), i.e., the on-device model is directly trained on all augmented data, the ROUGE-1 score also decreases from 0.391 to 0.360 on LaMP-2. 
Furthermore, when we fine-tune on-device models without LLM data augmentation (i.e., -LDA), the model performance further drops to 0.243 accuracy and 0.337 ROUGE-1 score on LaMP-2 and LaMP-7.
Overall, the results support the effectiveness of all the proposed components.

%%
% Hyperparameters
\begin{figure*}[t]
  \includegraphics[width=\textwidth]{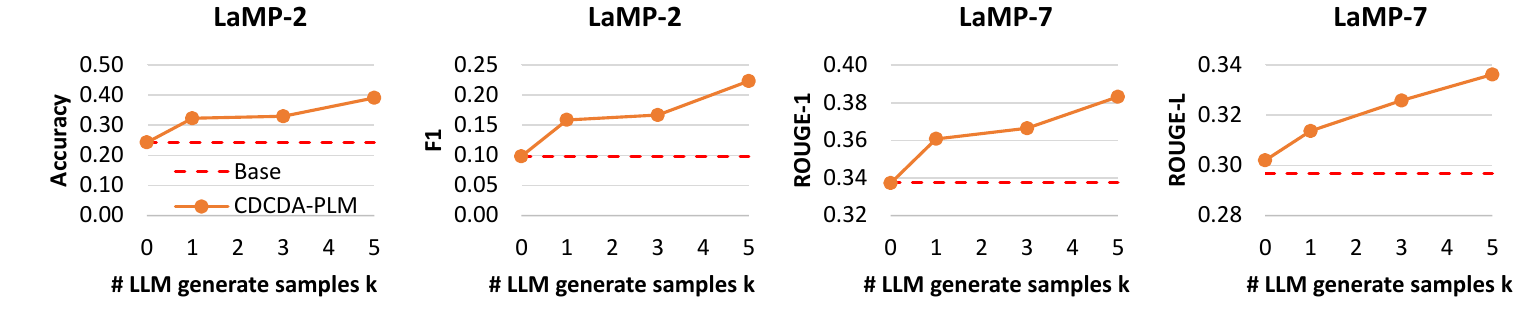}
  \Description{The performance increases as the number of LLM-generated samples increasing from 0 to 5.}
  \caption{The impact of hyperparameter in LLM data augmentation. $k$ controls the number of samples generated by server-sided LLM.}
  \label{fig:k samples generation}
\end{figure*}
\subsection{Hyper-parameter Analysis (RQ3)}
In this part, we investigate the impact of the hyper-parameters, synthetic data augmentation size $k$, associated with our proposed method. 
To better understand the impact of cloud-based LLM augmentation, we vary the number of LLM-generated samples 
$k$ for both the classification (LaMP-2) and generation (LaMP-7) tasks, as shown in Figure \ref{fig:k samples generation}. Overall, increasing 
$k$ leads to improvements in both tasks LaMP-2 and LaMP-7. Specifically, in both tasks, performance stabilizes on generating 1 and 3 samples and achieves greater improvement when increasing the generated samples to 5.

%% latency
\subsection{Efficiency Analysis (RQ4)}
% \caption{Inference efficiency results. Memory is the peak runtime memory, reported in MegaBytes (MiB). TTFT represents the time to first token and decode represents the average number of output tokens generated per second.}

% \usepackage{multirow}
% \usepackage{graphicx}
\begin{table}[t]
\centering
\caption{Inference efficiency results. Storage size is the required memory for deploying models. TTFT represents the time to generate the first token (sec), and Decode represents the average number of output tokens generated per second (tokens/s). Lower Storage Size and TTFT indicate better efficiency, whereas higher Decode speed is better.}
\label{tab:efficiency}
\begin{tabular}{l|cccccc}
\hline
\multicolumn{1}{c|}{\multirow{2}{*}{\textbf{Models}}} & \multirow{2}{*}{\textbf{\begin{tabular}[c]{@{}c@{}}Storage \\ Size (GB)\end{tabular}}} & \multicolumn{2}{c}{\textbf{Workstation}} &  & \multicolumn{2}{c}{\textbf{Android}} \\ \cline{3-4} \cline{6-7} 
\multicolumn{1}{c|}{} &  & TTFT & Decode &  & TTFT & Decode \\ \hline
\textbf{Cloud-based LLM} & 2.36 & 4.8 & 234 &  & 31.7 & 2.3 \\
\textbf{On-device LM} & 0.42 & 0.9 & 688 &  & 4.9 & 8.6 \\
 & (-82\%) & (-81\%) & (2.9x) &  & (-85\%) & (3.8x) \\ \hline
\end{tabular}%
\end{table}
To further evaluate the \modelname's efficiency on real-world edge devices, we deploy personalized on-device LM on two devices: (1) a workstation with GPU NVIDIA RTX A5000 GPU and (2) a Samsung Galaxy Tab A8. 
We first train all the on-device personalized LMs on a workstation with a GPU, and then they are encapsulated by MLC-LLM \citep{mlc-llm}, which is a compiler and high-performance deployment engine for LLMs. 
To compare the efficiency in on-device deployment, we evaluate the on-device model deployment's efficiency from two perspectives: model storage memory and inference efficiency (TTFT and decoding speed).
We record the inference efficiency for 5 runs of each model on the randomly selected queries from 6 tasks. 

The results are reported in Table \ref{tab:efficiency}. 
Our on-device models are much smaller and faster than the cloud-based models on two devices. 
Specifically, the memory footprint of our on-device model is more than five times smaller than that of the server-based model. 
The time-to-first-token (TTFT) for our on-device model is just 0.9 seconds on a workstation and 4.9 seconds on an Android phone—both lower than the TTFT of the server model. 
In terms of throughput, our on-device model generates output tokens at a rate three times faster than the original server model. 
Notably, according to \citet{humanReadSpeed}, the average human reading speed is approximately 4 to 6 words per second. 
Our on-device model, even on a low-end Android device, achieves a decoding speed of 8.6 tokens per second, which is sufficient for real-world applications.
Overall, the efficiency analysis proves that our developed personalized LM is able to achieve more practical memory and response latency for on-device services.

%% model comparison (T5)
\subsection{Generalization of Framework (RQ5)}
% \usepackage{multirow}
% \usepackage{graphicx}
\begin{table}[t]
\centering
\caption{Comparative results with FlanT5-base model on 6 tasks. In each task, the best result is marked in \textbf{bold}.}
\label{tab:model_compare}
\begin{tabular}{lccl|c}
\hline
\multicolumn{1}{c}{\multirow{2}{*}{\textbf{Tasks}}} & \multicolumn{2}{c}{FlanT5-base} &  & Qwen2.5-0.5B \\ \cline{2-3} \cline{5-5} 
\multicolumn{1}{c}{} & +BM25 & CDCDA-PLM &  & CDCDA-PLM \\ \hline
\textbf{LaMP-1 (Acc)} & \textbf{0.6} & 0.570 &  & 0.530 \\
\textbf{LaMP-2 (Acc)} & 0.3499 & 0.283 &  & \textbf{0.391} \\
\textbf{LaMP-3 (MAE)} & 0.4632 & 0.484 &  & \textbf{0.400} \\
\textbf{LaMP-4 (R-1)} & 0.0834 & 0.115 &  & \textbf{0.120} \\
\textbf{LaMP-5 (R-1)} & 0.2867 & \textbf{0.406} &  & 0.382 \\
\textbf{LaMP-7 (R-1)} & 0.2933 & 0.362 &  & \textbf{0.383} \\ \hline
\end{tabular}%
\end{table}
In this part, we investigate the effectiveness of our framework on another language model, FlanT5-base \citep{flant5_base}. 
As shown in Table \ref{tab:model_compare}, FlanT5-base with CDCDA-PLM achieves better performance than FlanT5-base with BM25 RAG on 4 tasks due to limited historical data on a single user, indicating that our framework is able to be implemented on different LM architectures to improve the on-device personalized performance.

%% personalization in synthetic samples
\begin{figure*}[t]
  \includegraphics[width=0.8\textwidth]{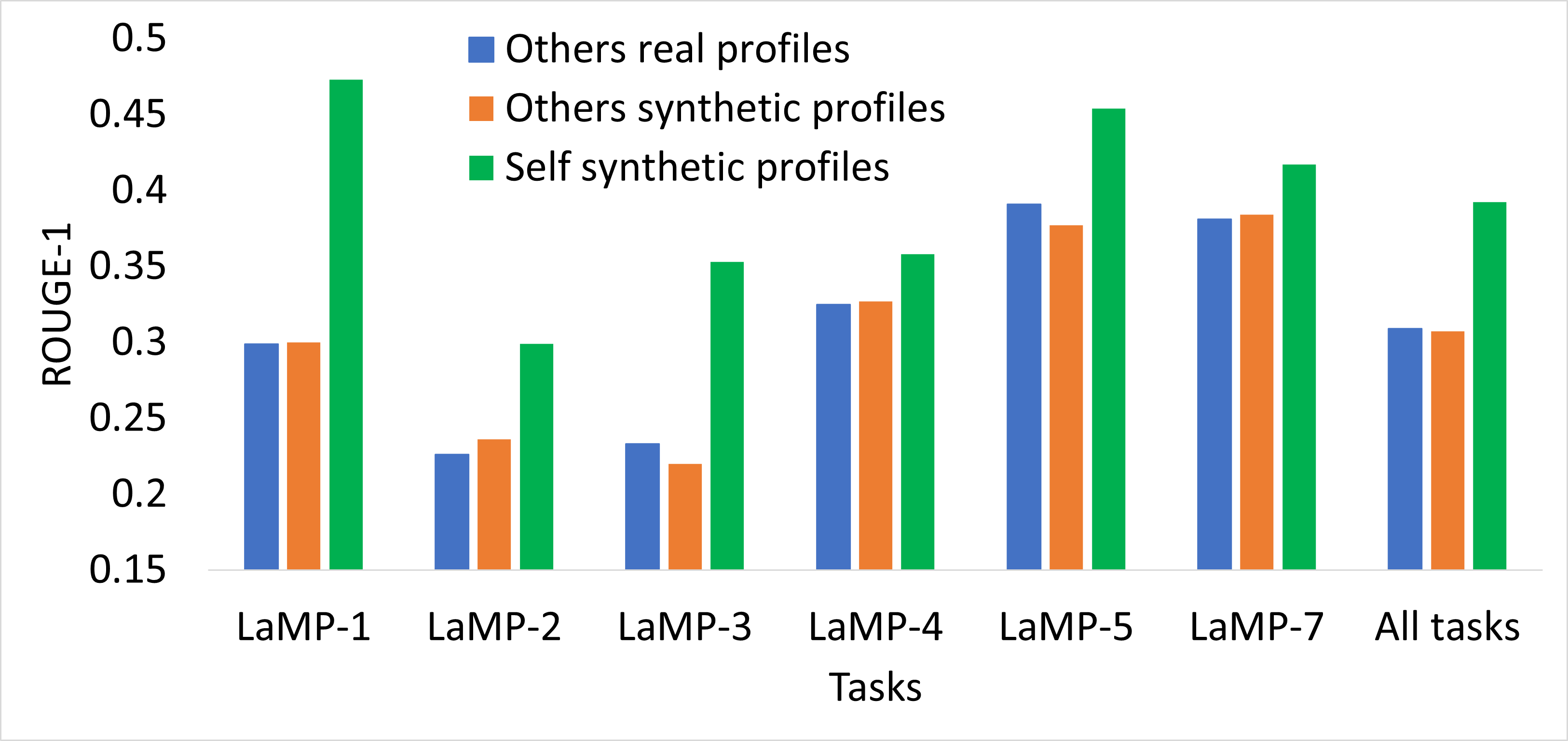}
  \Description{On all 6 tasks, a user's real profile is more similar to their own synthetic profile.}
  \caption{The comparison of the average ROUGE-1 score between each user's real profile and other users' real and synthetic profiles. ``All tasks'' reports the average on 6 tasks.}
  \label{fig:synthetic alignment}
\end{figure*}

\subsection{Personalization in Synthetic Data (RQ6)}
To ensure the alignment between synthetic data and user history, we conduct a quantitative analysis to assess whether the server-side LLM can generate effective personalized synthetic data.
Specifically, inspired by prior work \cite{richardson2023summaryretrieval}, we use OpenAI GPT-5-nano \cite{openai_gpt5_nano_2025} to generate a profile summarizing a user’s preferences based on their historical data. 
This is treated as the user’s real profile. 
We then also use GPT-5-nano to generate a synthetic user profile based on the user's synthetic data. 

To evaluate alignment, we compute ROUGE-1 between each user’s real profile and (1) their own synthetic profile, (2) other users’ real profiles, and (3) other users’ synthetic profiles. As reported in Figure~\ref {fig:synthetic alignment}, a user’s real profile exhibits higher similarity to synthetic counterparts than to other users' real profiles and synthetic profiles. 
These results support our claim that the LLM-generated data reflects user personalized preferences rather than generic or irrelevant content.

%% Case study
\subsection{Case Study}
\begin{figure*}[t]
  \includegraphics[width=\textwidth]{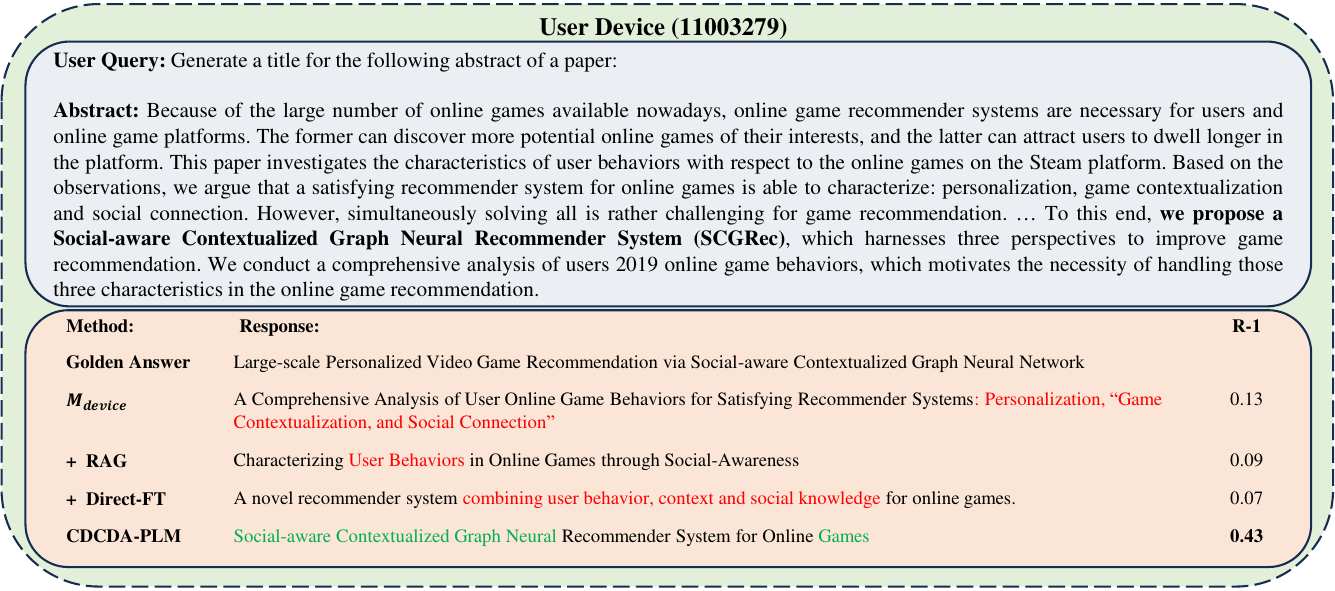}
  \Description{A case study for user 11003279. RAG personalized achieves 0.09 ROUGE-1 score. Direct fine-tune achieves 0.07. And our method achieves 0.43.}
  \caption{A case study in LaMP-5, which is the task of Personalized Scholarly Title Generation.}
  \label{fig:case study}
\end{figure*}

To further intuitively understand the personalization effectiveness of \modelname, we conduct a case study for a user on the Personalized Scholarly Title Generation (LaMP-5) task, which tests the ability of models to capture stylistic patterns when generating scholarly titles based on the abstract of an article.

% Case study for 11003279
Figure \ref{fig:case study} presents an example of a specific user. Note that, according to this user's historical data, they prefer to directly include the proposed method's name from the abstract as part of the title. In this case, they favor using the bold text ``Social-aware Contextualized Graph Neural Recommender System'' as indicated in the Golden Answer. However, all baseline models fail to capture this preference and instead generate titles by summarizing the abstract's semantics. Only our \modelname successfully identifies this pattern, producing a title most similar to the Golden Answer.

% \section{Conclusion}
\section{Limitations and Ethical Considerations}
Several limitations are concerned with our work.
% task specific personalization
Firstly, due to dataset constraints, our study aims to deploy a personalized model to generate responses on a specific task for each user, ignoring the user behaviors from other tasks and domains.
For example, for the user who engages in news headline generation and scholarly title generation tasks, both tasks could provide the user's stylistic pattern preference.
Moreover, in the future, we believe \modelname can be applied to any NLP task across different domains.
Secondly, the data quality of LLM augmentation can be affected by the cloud-based LLM. Exploring a larger LLM or multiple LLMs to augment user data remains an area for future investigation.
Thirdly, although CDCDA-PLM relies on the cloud LLM only during the synthetic data generation stage, it may limit applicability to the users in offline-first or highly restricted settings.
To address this challenge, our framework can be extended to employ a lightweight local LLM as on-device synthetic data generator. 

CDCDA-PLM uploading user data to server LLM for cloud-based data augmentation may leads to privacy concerns.
Therefore, our framework assumes that user data can be utilized under appropriate legal and ethical agreements, as is common in many real-world personalization systems (e.g., recommender services or cloud-based assistants). These agreements typically allow data collection and processing for the explicit purpose of improving user experience, often with user consent and privacy safeguards in place.
Nevertheless, CDCDA-PLM can be extended with additional privacy-preserving strategies to further reduce potential risks.
First, data minimization can be applied by transmitting only compact user profile summaries instead of raw text, thereby preventing the server from directly accessing sensitive content.
Secondly, differential privacy (DP) can be incorporated during user profile uploading by adding controlled noise, which prevents the server from reconstructing user history.
Therefore, it is important to investigate further robust methods for privacy protection in cloud-server LLM data augmentation. 

In addition, a personalized model aims to generate content aligning with user preferences and interests shown in user data. 
However, personalization models may be trained with user data consisting of biased and unfair information, leading to harmful responses. 
Within CDCDA-PLM, the biased data is uploaded to server LLM for augmentation, which further negatively affects the on-device model.
Future works may explore strategies to avoid sharing or augmenting harmful data on the server LLM.

\section{Conclusion}
This paper introduces a cloud-device collaborative data augmentation on-device personalized LM, named CDCDA-PLM, a personalized on-device LM deployment framework designed to close the performance and efficiency gap between cloud-based LLM and on-device LM by augmenting user historical data. 
Specifically, CDCDA-PLM first uses a server LLM to construct a synthetic dataset containing similar samples as user data to assist the on-device personalized model's fine-tuning.
The experimental results demonstrate that \modelname achieves better performance on personalized content generation.

\begin{acks}
The Australian Research Council partially supports this work under the streams of Future Fellowship (Grant No. FT210100624),  the Discovery Project (Grant No. DP240101108 and DP260100326), and the Linkage Projects (Grant No. LP230200892 and LP240200546).
\end{acks}

\bibliographystyle{ACM-Reference-Format}
\bibliography{A_MY_reference}

\appendix
\section{Prompt Details}

In this part, we show the prompt used in our experiment for Qwen2.5-Instruct.

\subsection{LaMP-1: Personalized Citation Identification}
<|im\_start|>system \\
With the given examples, which reference is related?<|im\_end|> \\
<|im\_start|>user \\
\{RETRIEVED USER HISTORY\} \\
For an author who has written the paper with the title \{PAPER TITLE\}, which reference is related? Just answer with [1] or [2] without explanation. [1]: \{OPTION\_1\} [2]: \{OPTION\_2\}<|im\_end|>

\subsection{LaMP-2: Personalized Movie Tagging}
<|im\_start|>system \\
With the given examples, generate a tag for the given movie.
<|im\_end|>\\
<|im\_start|>user \\
\{RETRIEVED USER HISTORY\} \\
Which tag does this movie relate to among the following tags? Just answer with the tag name without further explanation. tags: [sci-fi, based on a book, comedy, action, twist ending, dystopia, dark comedy, classic, psychology, fantasy, romance, thought-provoking, social commentary, violence, true story] description: \{MOVIE DESCRIPTION\}
<|im\_end|>

\subsection{LaMP-3: Personalized Product Rating}
<|im\_start|>system \\
With the given examples, generate a score for the given review.
<|im\_end|> \\
<|im\_start|>user \\
\{RETRIEVED USER HISTORY\} \\
What is the score of the following review on a scale of 1 to 5? just answer with 1, 2, 3, 4, or 5 without further explanation. review: \{REVIEW\}
<|im\_end|>

\subsection{LaMP-4: Personalized News Headline Generation}
<|im\_start|>system \\
With the given examples, generate a title for the given article. Only output the title and nothing else.
<|im\_end|> \\
<|im\_start|>user \\
\{RETRIEVED USER HISTORY\} \\
Generate a headline for the following article: \{ARTICLE\}
<|im\_end|>

\subsection{LaMP-5: Personalized Scholarly Title Generation}
<|im\_start|>system \\
With the given examples, generate a title for the given article. Only output the title and nothing else.
<|im\_end|> \\
<|im\_start|>user \\
\{RETRIEVED USER HISTORY\} \\
Generate a title for the following abstract of a paper: \{PAPER\}
<|im\_end|>

\subsection{LaMP-7: Personalized Tweet Paraphrasing}
<|im\_start|>system \\
With the given examples, paraphrase the following tweet without any explanation before or after it.
<|im\_end|> \\
<|im\_start|>user \\
\{RETRIEVED USER HISTORY\} \\
 Paraphrase the following tweet without any explanation before or after it: {USER TWEET}
<|im\_end|>

\section{Equations of Evaluation Metrics}
\label{metric_equation}
Accuracy measures the proportion of correct predictions among the total number of cases processed:
\begin{equation}
    \text{Accuracy} = \frac{TP + TN}{TP + TN + FP + FN},
\end{equation}
where \(TP\), \(FP\), \(TN\) and \(FN\) represent true positive, false positive, true negative, and false negative, respectively.
The F1-score is the harmonic mean of precision and recall:
\begin{equation}
    \text{F1-score} =2 \times \frac{precision \times recall}{precision + recall} = \frac{2 \times TP}{2 \times TP + FP + FN}.
\end{equation}
The MAE and RMSE measure the difference between predicted and reference values.:
\begin{equation}
    \text{MAE} = \frac{\sum_{i=1}^{n} |y_i - \hat{y}_i|}{n},
    \text{RMSE} = \sqrt{\frac{\sum_{i=1}^{n} (y_i - \hat{y}_i)^2}{n}},
\end{equation}
where \(n\) represents the number of predicted data. \(y_i\) is the ground-truth label of \(i\)-th sample and \(\hat{y}_i\) is the predicted value of \(i\)-th sample.

\end{document}